\documentclass[a4paper]{article}
\usepackage{epsfig}

\newcommand{\tensor}{\otimes}

\newcommand{\1}{\mathbf{1}}
\newcommand{\0}{\mathbf{0}}
\newcommand{\unit}{\mathbf{1}}
\newcommand{\re}{\mathbbm{R} \mbox{e }}
\newcommand{\im}{\mathbbm{I} \mbox{m }}
\newcommand{\C}{\mathbbm{C}}
\newcommand{\R}{\mathbbm{R}}

\usepackage{amsfonts}
\usepackage{bbm}

\begin{document}

\title{A Bayesian Analogue of Gleason's Theorem}\author{Thomas Marlow\\ \emph{School of Mathematical Sciences, University of Nottingham,}\\
\emph{UK, NG7 2RD}}

\maketitle

\begin{abstract}
We introduce a novel notion of probability within quantum history theories and give a Gleasonesque proof for these assignments.  This involves introducing a tentative novel axiom of probability.  We also discuss how we are to interpret these generalised probabilities as partially ordered notions of preference and we introduce a tentative generalised notion of Shannon entropy.  A Bayesian approach to probability theory is adopted throughout, thus the axioms we use will be minimal criteria of rationality rather than \emph{ad hoc} mathematical axioms.
\end{abstract}



\section{Introduction and Summary}

In \cite{Marlow06complex} we postulated a novel notion of probability by generalising Cox's axioms of probability \cite{Cox46,CoxBOOK} in a manner appropriate to quantum theory.  In this paper we wish to go one step further; we will present a uniqueness proof analogous to Gleason's theorem for our postulated generalised probabilities.  We will be helped considerably by another analogue of Gleason's theorem in the literature \cite{ILS94} which is applied to the decoherence functional in the History Projection Operator (HPO) formulation of the consistent histories programme \cite{Isham94}.  First we will review results previously discussed \cite{Marlow06complex} and then we will outline the relevant Gleason-like theorem, its interpretation and, for completeness, its proof.  We will then propose a generalised entropy.

We will adopt a Bayesian approach to probability theory and we will use Cox's approach in particular \cite{CoxBOOK}.  Probabilities are usually considered real numbers because of an association with relative frequencies.  As soon as we adopt an approach to probability theory where we merely assign probabilities as an ordered notion of preference then there is absolutely no \emph{a priori} reason to consider probabilities as real numbers.  One might try to design `zeroth' axioms of probability theory which end up ensuring that probabilities are in fact real numbers, and then one might introduce further axioms to constrain how we assign these real numbers to propositions.  Such an approach is, however, problematic because such `zeroth' axioms are rather \emph{ad hoc}.

Consider some arbitrary propositions $\alpha, \beta$ and $\gamma$ to which we are to assign probabilities.  Consider also a notion of ordering `$>$' to be defined on the probability space.  Two possible `zeroth' axioms \cite{JaynesBOOK}, which constrain how this ordering notion is to behave, are `universal transitivity':

\begin{eqnarray}
\mbox{Axiom 0a:}& \mbox{ If } p(\alpha \vert I) > p(\beta \vert I) \mbox{ and } p(\beta \vert I) > p(\gamma \vert I) \nonumber \\ & \mbox{ then } p(\alpha \vert I) > p(\gamma \vert I),
\end{eqnarray}

\noindent and `universal comparability':

\begin{eqnarray}
\mbox{Axiom 0b:}& \mbox{ For all } \alpha, \beta \mbox{ we have that either } p(\alpha \vert I) > p(\beta \vert I) \nonumber \\ & \mbox{ or } p(\alpha \vert I) < p(\beta \vert I) \mbox{ or }p(\alpha \vert I) = p(\beta \vert I).
\end{eqnarray}

Given these zeroth axioms it would seem natural (although it is still not strictly necessary) to use real numbers for probability assignments.  Axiom 0a is often considered desirable because probabilities are intended to represent transitive notions of preference in some sense.  Axiom 0b is, however, far more dubious and there is a history in the literature of people trying not to assume it (see Appendix A.3 of \cite{JaynesBOOK} for references and also see \cite{Youssef94} and \cite{Isham02}).  Why presume that we can probabilistically compare all propositions universally, especially in quantum theory where some propositions are considered `incompatible' or `complimentary'?  It is prudent to not assume axiom 0b from the outset (it might be that we are forced to adopt it later).  If we were to adopt axiom 0b then it might be that we will be introducing relationships between probabilities that we are not justified in invoking---and any problems like nonadditivity and so forth might be due to such a mistaken assumption.

Rather than adopt these two controversial zeroth axioms let us, for the time being, use a weaker zeroth axiom that we can all surely agree on:

\begin{equation}
\mbox{Axiom 0}': \mbox{ If } \alpha \leq \beta \mbox{ then, presumably, } p(\alpha \vert I) \leq p(\beta \vert I),
\label{monoticity}
\end{equation}

\noindent where `$\leq$' is, in the least, a partial order in the context of both the proposition space and the space of generalised probabilities\footnote{In the context of standard probability theory we could use Venn diagrams to define $\alpha \leq \beta$ using subset inclusion and we would define $p(\alpha \vert I) \leq p(\beta \vert I)$ using the total ordering of real numbers.}.  We call this axiom `monoticity' \cite{IB98}.  So as to avoid confusion with standard probability theory (theories which obey axioms 0a and 0b) it is prudent to call any assignment which merely obeys the weaker zeroth axiom (\ref{monoticity}) by another name: we will call them `pedagogical examples' or `pegs'.  Probabilities are then special examples of pegs.

Our task in this paper is then to find a peg theory for a histories algebra.  We will use the histories propositional algebra $\cal{P(V)}$, where ${\cal V} = \tensor^{n} \cal{H}$, which was originally introduced by Isham \cite{Isham94}.  The natural connectives on this space of projection operators are the standard $\wedge, \vee, \neg$ connectives and we use the standard partial order $\leq$ upon it \cite{Isham94}.  A homogeneous history proposition $\alpha$ is defined as a time ordered tensor product of projection operators $\hat{\alpha}_{t_m} \in \cal{P(H)}$:

\begin{equation}
\alpha := \hat{\alpha}_{t_n} (t_n) \tensor \hat{\alpha}_{t_{n-1}}(t_{n-1}) \tensor ... \tensor \hat{\alpha}_{t_1}(t_1).
\end{equation}

\noindent  We stay in the Heisenberg picture such that each projection operator has the dynamics already implicit such that $\hat{\alpha}_{t_m} (t_m) = \hat{U}^\dagger(t_m - t_{m-1}) \hat{\alpha}_{t_m} \hat{U}(t_m - t_{m-1})$ where $\hat{\alpha}_{t_m}$ are Schr\"odinger picture operators.  Our novel peg assignments, those that we suggested in \cite{Marlow06complex}, are

\begin{equation}
p(\alpha \vert I) := \mbox{tr}_{\cal{H}}(C_\alpha \hat{\rho})
\label{complexprob}
\end{equation}

\noindent where $C_\alpha = \hat{\alpha}_{t_n} (t_n) \hat{\alpha}_{t_{n-1}}(t_{n-1}) ... \hat{\alpha}_{t_1}(t_1)$ and $\hat{\rho}$ is a density operator on $\cal{H}$.  We keep an explicit hypothesis $I$ in the notation because such a thing is natural when discussing probabilities from a Bayesian perspective \cite{Mana04}, and it avoids us confusing peg assignments that are made given different prior information.  Clearly these pegs might obey (\ref{monoticity}) by relating the natural partial order on ${\cal P(V)}$ to some partial order on $\C$.  Once we introduce further axioms that these pegs ought to obey---other than (\ref{monoticity})---then we will be able to speculate what this partial order on the peg space might be.

We proposed these pegs for the history algebra ${\cal P(V)}$ because they are additive for disjoint homogeneous history propositions \cite{Marlow06complex}---thus these complex pegs seem to \emph{behave} something like we expect probabilities should.  Now we aim to show that we can \emph{derive} these pegs from axioms analogous to Cox's axioms of probability theory applied to the HPO algebra using an analogue of Gleason's theorem.

\section{A Gleason Analogue}

So, let us remind ourselves of what Gleason's theorem \cite{Gleason57} tells us.  Gleason's theorem is about trying to assign probabilities to a quantum propositional algebra.  In standard quantum theory the relevant propositional algebra is taken to be $\cal{P(H)}$, the set of projection operators upon a Hilbert space $\cal{H}$, where the natural logical connectives on $\cal{P(H)}$ are $\wedge, \vee, \neg$ and we denote the standard partial order relation `$\leq$'.  So, naturally, a probability assignment should obey certain rules which we shall use to define what is often, perhaps confusingly, called a state (we call it a state more because of what we end up proving).  A state $\sigma \in \cal{S}$ is a real valued function on $\cal{P(H)}$ which has the following properties:

\begin{enumerate}

\item \emph{Positivity}: $\sigma(\hat{P}) \geq 0$ for all $\hat{P} \in \cal{P(H)}$,

\item \emph{Additivity}: if $\hat{P}$ and $\hat{R}$ are disjoint---$\hat{P} \wedge \hat{R} = \hat{\0}$---then $\sigma(\hat{P} \vee \hat{R}) = \sigma(\hat{P}) + \sigma(\hat{R})$,

\item \emph{Normalisation}: $\sigma(\hat{\unit}) = 1$,

\end{enumerate}

\noindent where $\hat{\0} \in \cal{P(H)}$ is the proposition that is always false, and $\hat{\unit} \in \cal{P(H)}$ is the proposition that is always true.  Gleason's theorem is simply that states assigned to $\cal{P(H)}$, for dim$\cal{H} > \mbox{2}$, are in one-to-one correspondence with density matrices on $\cal{H}$ such that

\begin{equation}
\sigma_{\rho}(\hat{P}) = \mbox{tr}(\hat{P} \hat{\rho}) \mbox{ for all } \hat{P} \in \cal{P(H)}.
\end{equation}

\noindent  One takes the propositional algebra of projection operators, makes basic assumptions about how probabilities ought to behave, and one \emph{derives} that such probability assignments are in one-to-one correspondence with density matrices.  The axioms of probability theory ensure the density matrix structure of quantum theory.

In analogy with Gleason's theorem we should list a set of axioms that our pegs should obey and then try and derive the theorem from there.  For these axioms, we argue, we should look to Cox's axioms of probability theory which ensure that we don't introduce functional relationships between peg assignments that aren't rationally justified.  Cox's $\wedge$-axiom is simply that the pegs we peg to propositions conjoined using the `and' operation should be limited to be functionally dependent only on some very specific pegs:

\begin{equation}
p(\alpha \wedge \beta \vert I) := F[p(\alpha \vert \beta I), p(\beta \vert I)]
\label{COX1}
\end{equation}

\noindent where $F$ is an arbitrary function that is sufficiently well-behaved for our purposes.  Similarly, Cox's $\neg$-axiom is that the peg we peg to the negation of a proposition should only functionally depend upon the peg of the proposition before it was negated:

\begin{equation}
p(\neg \alpha \vert I) := G[p(\alpha \vert I)].
\label{COX2}
\end{equation}

\noindent These two axioms are criteria of rationality that are at the heart of Cox's approach to probability theory and these are all he needed (except for the additional assumption that probabilities are real numbers) in order to prove the basics of probability theory as applied to a Boolean algebra of propositions.

Cox's two axioms suggest we should use a peg that is additive for disjoint history propositions and that is normalised \cite{CoxBOOK}.  It turns out that this will not be sufficient for our peg theory; we will need a further axiom.  Luckily one is forthcoming.  Note that in the HPO formulation of history theories we have the three natural logical connectives $\wedge,\vee,\neg$ which correspond roughly to `and', `or' and `negation' operations (although with the standard non-distributivity issue we have in quantum theory).  Clearly, however, when going from projection operators defined at a single-time to explicitly history orientated propositions there is another natural connective, namely changing the temporal order.  As such we can define an operator $M$ which reverses the order on any tensor product vector, $M(v_1 \tensor v_2 \tensor ... v_m) := (v_m \tensor v_{m-1} \tensor ... v_1)$.  Thus the temporal reversal of the Heisenberg picture history proposition $\alpha$ is given by:

\begin{equation}
\lhd \alpha := M \alpha M = \hat{\alpha}_{t_1} (t_1) \tensor \hat{\alpha}_{t_{2}}(t_{2}) \tensor ... \tensor \hat{\alpha}_{t_n}(t_n).
\end{equation}

Note that $\lhd$ reverses both the kinematical and the dynamical temporal orderings because we are in the Heisenberg picture (\emph{cf.} \cite{SavvidTHESIS}).  Applying $\lhd$ twice in this manner obviously gives back the same history proposition; behaviour that is analogous to the `negation' operation.  Hence we should introduce a third peg axiom analogous to (\ref{COX2}):

\begin{equation}
p(\lhd \alpha  \vert I) := H[p(\alpha \vert I)]
\label{COX3}
\end{equation}

\noindent where $H$ is an arbitrary function that is sufficiently well-behaved for our purposes.  The peg we assign to such a proposition should be assigned in a noncontextual manner according to (\ref{COX3}).  Clearly, by Youssef's argument \cite{Youssef94}, we could use complex numbers and still keep consistency with analogues of Cox's two axioms.  Our tentative third axiom seems to give our pegs an `extra degree of freedom'.  Using real numbers would not be a possibility for a peg that also obeys (\ref{COX3}) and luckily we are using the uncontroversial weaker zeroth axiom (\ref{monoticity}).  Hence it might be that we can find a peg which obeys (\ref{monoticity}), (\ref{COX1}), (\ref{COX2}) and (\ref{COX3}).

In analogy with how we define states in Gleason's theorem let us define complex assignments as maps $l: \cal{P(V)} \rightarrow \C$ which obey the following rules:

\begin{enumerate}

\item \emph{Conjugation}.  $l(\alpha)^* = l(\lhd \alpha)$ for all $\alpha$,

\item \emph{Additivity}.  If $\alpha$ and $\beta$ are disjoint then $l(\alpha \vee \beta) = l(\alpha) + l(\beta)$,

\item \emph{Normalisation}.  $l(\unit) = 1$.

\end{enumerate}

\noindent  We use the similar notation as we used for the Hilbert space $\cal{H}$ because $\cal{V}$ is still a Hilbert space (although we leave the `hats' off operators in ${\cal V}$).  This is an advantage of using Isham's HPO formulation of the history algebra \cite{Isham94}.

Note that the peg axioms (\ref{COX1}), (\ref{COX2}) and (\ref{COX3}) do not uniquely ensure that we \emph{must} use the complex assignments $l$---just as Cox's axioms in standard probability theory don't ensure that we \emph{must} use real numbers \emph{per se}.  The peg axioms ensure that, whatever assignments we do use for convenience, such assignments at least obey the relevant criteria of rationality.  Hence we do not argue that the complex assignments $l$ are uniquely the only pegs we could use, but clearly the maps $l$ do obey (\ref{COX1}), (\ref{COX2}) and (\ref{COX3}) and might yet obey (\ref{monoticity}) for some partial order on $\C$.  One might even argue that it is not the particular assignments that matter; it is the catalogue of functional relationships between them that are important (these we categorise axiomatically using analogues of Cox's axioms).  Nonetheless, it is convenient to use particular representations (just as it is convenient to use real numbers in standard Bayesian probability theory).

Can we now start to tackle a proof of a Gleason-like theorem for these complex assignments $l$?  In fact, the result follows from an analogue of Gleason's theorem for decoherence functionals already in the literature \cite{ILS94}.

Let us first review some identities \cite{ILS94}.  Note that

\begin{equation}
\mbox{tr}_{\cal{H}}(\hat{A}_1 \hat{A}_2 ... \hat{A}_n) = \mbox{tr}_{\tensor^n \cal{H}}(\hat{A}_1 \tensor \hat{A}_2 \tensor ... \hat{A}_n S)
\label{identity}
\end{equation}

\noindent where $\hat{A}_m$ are arbitrary self adjoint operators on $\cal{H}$ and $S$ is a linear operator $S: \tensor^n \cal{H} \rightarrow$ $\tensor^n \cal{H}$ defined on product vectors by

\begin{equation}
S(v_1 \tensor v_2 \tensor ... \tensor v_n) := v_2 \tensor v_3 \tensor ... \tensor v_n \tensor v_1
\end{equation}

\noindent and extended by linearity and continuity to give a unitary operator on $\tensor^n \cal{H}$.

We can swap between the Heisenberg and Schr\"odinger pictures using:

\begin{equation}
\mbox{tr}_{\cal{H}}(C_\alpha) = \mbox{tr}_{\tensor^{n} \cal{H}}(\hat{\alpha}_{t_n} \tensor \hat{\alpha}_{t_{n-1}} \tensor ... \hat{\alpha}_{t_1} S^U).
\end{equation}

\noindent where the $\hat{\alpha}_{t_m}$ are Schr\"{o}dinger picture projection operators and $S^{U} := (\hat{U}^{\dagger}_1 \tensor...\tensor \hat{U}^{\dagger}_n)S(\hat{U}_1 \tensor...\tensor \hat{U}_n)$.

In an analogous way \cite{ILS94}, we can absorb the initial state into an operator defined on ${\cal V} = \tensor^n {\cal H}$ using the identity (\ref{identity}).  Note that

\begin{eqnarray}
&&\mbox{tr}_{\cal{H}}(\hat{A}_1 \hat{A}_2 ... \hat{A}_n)  \nonumber \\
&=& \mbox{tr}_{\tensor^n \cal{H}}((\hat{A}_1 \tensor \hat{A}_2 \tensor ... \hat{A}_{n-1} \tensor \unit_n)(\unit_1 \tensor \unit_2 \tensor ... \unit_{n-1} \tensor \hat{A}_n) S)\\
&=& \mbox{tr}_{\tensor^n {\cal H}}((\hat{A}_1 \tensor \hat{A}_2 \tensor ... \hat{A}_{n-1} \tensor \unit_n) Y)\\
&=& \mbox{tr}_{\tensor^{n-1} \cal{H}}((\hat{A}_1 \tensor \hat{A}_2 \tensor ... \hat{A}_{n-1} Y')
\end{eqnarray}

\noindent where $Y'$ is obtained from $Y$ by tracing over a complete set of states for the $n$th Hilbert space.  The form of (\ref{identity}) is preserved under removing an operator by the action of a partial tracing, and is also preserved when removing the dynamics from around each single-time proposition in the history proposition.

So, using these identities we can absorb all the dynamics and initial state into some operator $Z$ such that:

\begin{equation}
\mbox{tr}_{\cal{H}}(C_\alpha \hat{\rho}) = \mbox{tr}_{\tensor^{n} \cal{H}}(\hat{\alpha}_{t_n} \tensor \hat{\alpha}_{t_{n-1}} ... \tensor \hat{\alpha}_{t_1} Z_{\rho, H}).
\label{Z}
\end{equation}

\noindent  Note that the LHS of (\ref{Z}) is in the Heisenberg picture whereas the RHS is in the Schr\"odinger picture---we have split the dynamics and kinematics into distinct entities.  There are good reasons for doing this as it would allow us to investigate the distinction between the two forms of temporal orderings \cite{SavvidTHESIS}.  But note that we can stay within the Heisenberg picture if we wish (for it is, by far, the preferable picture \cite{Rovel02}); we keep the dynamics around the corresponding projection operators and absorb just the initial state into an operator $Y$:

\begin{equation}
\mbox{tr}_{\cal{H}}(C_\alpha \hat{\rho}) = \mbox{tr}_{\tensor^{n} \cal{H}}(\hat{\alpha}_{t_n}(t_n) \tensor \hat{\alpha}_{t_{n-1}}(t_{n-1}) \tensor ... \hat{\alpha}_{t_1}(t_1) Y_{\rho}).
\label{Y}
\end{equation}

The above ensures that we can put our tentative assignment into `Gleason' form.  Now we can prove an analogue of Gleason's theorem for such operators $Y_{\rho}$.  The theorem and proof follows the analysis in \cite{ILS94} almost word for word.  There are, however, distinguishing features and, for completeness, we repeat the analysis here since the proof is so short.

\subsection*{Theorem}

If dim ${\cal V} > 2$, the complex assignments $l$ are in one-to-one correspondence with operators $Y$ on $\cal{V}$ $= \tensor^n \cal{H}$ according to the rule

\begin{equation}
l(\alpha) = \mbox{tr}_{\cal{V}}(\alpha Y)
\label{Gleasonesque}
\end{equation}

\noindent with the restrictions that:

\begin{eqnarray}
&\mbox{a) }& Y^\dagger = MYM \label{a}\\
&\mbox{b) }& \mbox{tr}_{\cal{V}}(Y)  = 1 \label{b}
\end{eqnarray}

\noindent where $M$ is an operator that reverses the order of the entries in a tensor product vector; $M(v_1 \tensor v_2 \tensor ... v_m) := (v_m \tensor v_{m-1} \tensor ... v_1)$.

\subsection*{Proof}

In one direction, the theorem is trivial; if a function $l$ is \emph{defined} by the right hand side of (\ref{Gleasonesque}) it clearly obeys the crucial additivity condition.  The extra requirements (\ref{a}) and (\ref{b}) ensure normalisation and conjugation requirements.

Conversely, let $l: \cal{P(V)} \rightarrow \C$ be a complex assignment.  The proof that it must have the form (\ref{Gleasonesque}) exploits Gleason's theorem applied to $\cal{P(V)}$.

Let $\re l$ and $\im l$ denote the real and imaginary parts of $l$ so that

\begin{equation}
l(\alpha) = \re l(\alpha) + \im l(\alpha)
\end{equation}

\noindent where $\re l(\alpha) \in \R$ and $\im l(\alpha) \in \R$.  The additivity condition on $l$ means that $\re l(\alpha)$ and $\im l(\alpha)$ are additive functions on $\cal{P(V)}$, \emph{i.e.}, $\re l(\alpha_1 \vee \alpha_2) = \re l(\alpha_1) + \re l(\alpha_2)$ for any disjoint pair $\alpha_1,\alpha_2$ of projectors and similarly for $\im l$.  We have that $l(\alpha)$ is a continuous function of its argument and hence $\alpha \mapsto l(\alpha)$ is a continuous function on $\cal{P(V)}$, as are its real and imaginary parts.  However the set of all projectors in the finite dimensional space $\cal{V}$ is a finite disjoint union of Grassman manifolds and is hence compact.  It follows that the functions $\alpha \mapsto \re l(\alpha)$ and $\alpha \mapsto \im l(\alpha)$ are bounded below and above.  On the other hand, for any $r \in \R$, the quantity

\begin{equation}
\kappa_r(\alpha) := r \mbox{dim}(\alpha) = r \mbox{tr}(\alpha)
\end{equation}

\noindent is a real additive function of $\alpha$, and hence so are $\re l + \kappa_r$ and $\im l + \kappa_s$ for any $r,s \in \R$.  We can choose an $r$ such that $\re l + \kappa_r \geq 0$ for all $\alpha$ (and $s$ such that $\im l + \kappa_s \geq 0$) and due to an upper bound we can choose positive real scale factors $\mu$ and $\nu$ such that for all $\alpha$ we have that

\begin{eqnarray}
0 \leq \mu(\re l + \kappa_r)(\alpha) \leq 1\\
0 \leq \nu(\im l + \kappa_s)(\alpha) \leq 1.
\end{eqnarray}

These inequalities plus the additivity property show that, for each $\alpha \in \cal{P(V)}$, the quantities $\alpha \mapsto \mu(\re l + \kappa_r)(\alpha)$ and $\alpha \mapsto \nu(\im l + \kappa_s)(\alpha)$ are states on the lattice $\cal{P(V)}$.  Then Gleason's theorem shows that there exists a pair of density matrices $\rho^1$ and $\rho^2$ on $\cal{V}$ such that for all $\alpha \in \cal{P(V)}$,

\begin{eqnarray}
\mu(\re l + \kappa_r)(\alpha) = \mbox{tr}_{\cal{V}}(\rho^1 \alpha) \\
\nu(\im l + \kappa_s)(\alpha) = \mbox{tr}_{\cal{V}}(\rho^2 \alpha)
\end{eqnarray}

\noindent and so

\begin{eqnarray}
\re l(\alpha) = \mbox{tr}_{\cal{V}}((\frac{1}{\mu}\rho^1 - r)\alpha) = \mbox{tr}_{\cal{V}}(Y^1 \alpha) \\
\im l(\alpha) = \mbox{tr}_{\cal{V}}((\frac{1}{\nu}\rho^2 - s)\alpha) = \mbox{tr}_{\cal{V}}(Y^2 \alpha)
\end{eqnarray}

\noindent where $Y^1:=\frac{1}{\mu}\rho^1 - r$ and $Y^2:=\frac{1}{\nu}\rho^2 - s$.  Thus we have shown the existence of a family of operators $Y:=Y^1 + i Y^2$ on $\cal{V}$ such that

\begin{equation}
l_{\rho}(\alpha) = \mbox{tr}_{\cal{V}}(\alpha Y_{\rho})
\label{unique}
\end{equation}

This completes the proof of the theorem because the conditions (\ref{a}) and (\ref{b}) follow at once from the conjugation and normalisation conditions on complex assignments. $\square$ \\

We do not discuss any extensions to infinite dimensional cases.  We add the subscript $\rho$ to $Y$ to emphasise that it depends upon the initial state; it is anticipated that $Y_{\rho}$ can be decomposed into some operators on $\cal{V}$ which are universally defined (through relations between traces of products of operators and traces of tensor product operators) and some operators that are related to the initial state.  Clearly the $Y$ operators on ${\cal V}$ and density operators on $\cal{H}$ are intimately related, the task is now to investigate the properties and interpretation of these pegs.  But, in the least, we can put our assignment (\ref{complexprob}) into the form (\ref{unique}) for which we have an analogue of Gleason's theorem.

One issue that we have to identify is that we have promoted the operation $\lhd$ to a connective on par with $\vee, \wedge, \neg$, and it may not seem natural to some to do this.  We considered it natural because we were going from a space ${\cal P(H)}$ which identified propositions at a single time to a space ${\cal P(V)}$ which explicitly identified history propositions.  So we need a connective that can relate different temporal orderings.  One might then query, why specifically $\lhd$?  Why not some other operation, like making any permutation of single-time entries?  Staying in the Heisenberg picture ensures that the dynamics are already taken care of, and permuting the entries would mess up this fact---hence we only discuss a connective that maintains the dynamical relationships between single-time propositions.

Note that we do not need to use the monoticity axiom (\ref{monoticity}) for Gleason-like proofs; it is, in some sense, redundant. Nonetheless, how might our pegs obey monoticity?   In ${\cal P(V)}$ we have the following condition:

\begin{equation}
 \0 \leq \alpha \leq \1 \mbox{ for all } \alpha \in {\cal P(V)}.
\end{equation}

\noindent  For our pegs we have that $p(\0 \vert I) = 0$ and $P(\1 \vert I) = 1$ and hence, by monoticity, we must, in the least, demand that

\begin{equation}
0 \leq p(\alpha \vert I) \leq 1 \mbox{ for all } \alpha \in {\cal P(V)}.
\label{constraint}
\end{equation}

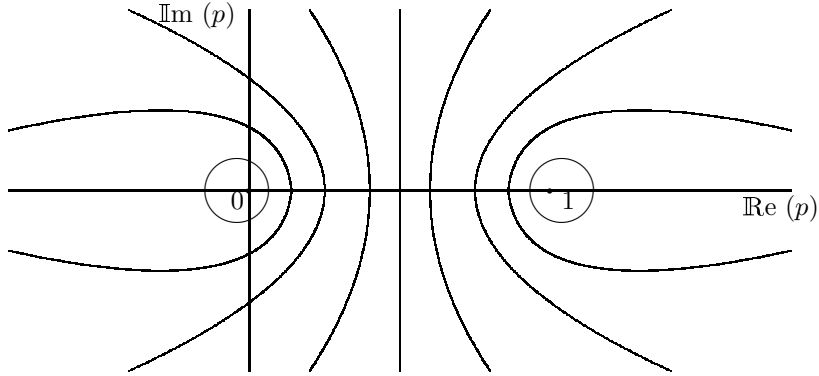
\begin{figure}[h!]
\unitlength=0.80mm
\begin{picture}(130.00,60.00)


\bezier{500}(00.00,30.00)(50.00,30.00)(130.00,30.00)%
\bezier{500}(40.00,00.00)(40.00,30.00)(40.00,60.00)%
\put(25,58){{$\im(p)$}}
\put(122,26){{$\re(p)$}}
\put(37,27){0}
\put(92,27){1}
\put(90,30){\circle*{1}}
\put(40,30){\circle*{1}}


\bezier{500}(65.00,00.00)(65.00,30.00)(65.00,60.00)%

\bezier{500}(50.00,00.00)(70.00,30.00)(50.00,60.00)
\bezier{500}(20.00,00.00)(85.00,30.00)(20.00,60.00)%
\bezier{500}(00.00,40.00)(45.00,50.00)(47.00,30.00)%
\bezier{500}(00.00,20.00)(45.00,10.00)(47.00,30.00)%
\put(38,30){\circle{10}}

\bezier{500}(80.00,00.00)(60.00,30.00)(80.00,60.00)
\bezier{500}(110.00,00.00)(45.00,30.00)(110.00,60.00)%
\bezier{500}(130.00,40.00)(85.00,50.00)(83.00,30.00)%
\bezier{500}(130.00,20.00)(85.00,10.00)(83.00,30.00)%
\put(92,30){\circle{10}}

\end{picture}
\begin{quote}
\caption{\label{partial1} Contour lines of pegs of equivalent size by a suggested partial order on $\C$.  The `height' of each contour gets larger starting from the circle around $0$ until we reach the circle around $1$.}
\end{quote}
\end{figure}

\noindent  One tentative partial order might look something like Fig.(\ref{partial1}).  This partial order\footnote{Equivalently, one can picture the same partial order as allowing many different paths from uncertainty to certainty such that these paths are symmetric in the real axis.   These paths would look like the lines of magnetic flux between North (0) and South (1) magnetic poles in 2D (not illustrated).  This absolves us of Jaynes' argument against comparative probability theories---ones that don't obey axiom 0b---by allowing many dense paths from $0$ to $1$, see the appendix of \cite{JaynesBOOK}).} has the added advantage as we are unable to relate $p(\lhd \alpha \vert I)$ and $p(\alpha \vert I)$ using it (the partial ordering is symmetric in the real axis and complex conjugation represents time reversal).  Since, by the partial order on ${\cal P(V)}$, $\alpha$ NR $\lhd \alpha$ (where NR stands for `not related to by the relevant partial order') we ought to demand this of the peg space as well, such that $p(\lhd \alpha \vert I)$ NR $p(\alpha \vert I)$.  Also, this partial order reduces to the standard probabilities when we move to the real line between $0$ and $1$.

Note, however, that there are many partial orders on $\C$ and there might be another applicable one---we introduce the partial order represented by Fig.(\ref{partial1}) to emphasise that monoticity will provide constraints for the partial orders that we can use.  For disjoint propositions $\alpha$ and $\beta$ we have that $\alpha \leq \neg \beta$ hence the monoticity requirement ensures that $p(\alpha \vert I) \leq 1 - p(\beta \vert I)$ which is satisfied as long as (\ref{constraint}) is.  The constraint (\ref{constraint}) thus ensures a class of partial orders that might be useful and Fig.(\ref{partial1}) seems the most apt.

Even obeying all the peg axioms we have to hand it is still very difficult to interpret these pegs as `probabilities' \emph{per se} because we are taught again and again that probabilities are real numbers.  However, remember that there is no \emph{a priori} reason to regard probabilities as real numbers, they are merely magnitudes that we use in order to assign a partially ordered notion of preference to propositions.

\section{Discussion}

Having an analogue of Gleason's theorem for our pegs is not enough;  we now need to argue why we should use such complex pegs in the first place.  We have given an argument based on rejecting axiom 0b so now it is important to discuss the factitious problems that are solved by rejecting it.  If we keep axiom 0b then, by the very fact that we will be comparing propositions in a manner that is not justified rationally, we will be introducing relations within the probability space that are not underpinned by relations in the proposition space.  Hence we call such relations in the probability space `factitious'.

The language we have used here is perhaps quite telling.  Our use of the term `factitious' harks back to Einstein's use of the term.  Clearly, if we adopt a relational approach to theory building (one which obeys Leibniz's principles of relationalism \cite{Smolin05}) we do not want to introduce factitious elements in any theory.  Cox adopted a rationalist approach analogous to relationalism---he gave a formulation of probability theory which ensures that factitious functional relations between propositions, quantified using pegs, are never introduced.  Instead we only maintain those functional relationships that we can justify.  Clearly this peg approach might therefore be useful in the quantum gravity domain where an explicitly relational approach is often considered a requirement.  One might na\"{\i}vely argue that, in addition to a relational notion of spacetime, one needs a relational notion of probability.  In the least it would be prudent to adopt an approach to probability theory based upon criteria of rationality rather than \emph{ad hoc} axioms.

Hence it is plausible that problems like nonlocality are factitious problems that are caused by invoking a probability theory which does not have sufficient structure.  Thus a Bayesian approach may bring locality back to quantum theory, just as Einstein bought locality back to gravitational theories in building general relativity \cite{Gisin05}---in fact this is the major reason many invoke Bayesian reasoning within quantum theory \cite{Fuchs02} (although one does not need to adopt as drastic a peg theory as we adopt here in order to tentatively deny nonlocality).

Similarly, the problem of hidden variables might also be factitious.  The Kochen-Specker theorem \cite{KS} seems \cite{Apple04} to prove that we cannot assign definitive values to variables prior to measurement in quantum theory.  This is obviously compatible with a Bayesian approach to ignorance---we cannot assign such values because we are explicitly \emph{presuming we are ignorant of them}.

These complex pegs are intimately related to the approaches of Feynman \cite{Feynman87} and Hartle \cite{Hartle04} who invoke real `probabilities' that can lie outside of $[0,1]$.  Hartle's virtual `probabilities' are explicitly found using the real parts of our complex pegs (\ref{complexprob}) which, in turn, were originally introduced by Goldstein and Page \cite{GP95} in their linearly positive histories approach.  Thus the linearly positive histories and the consistent histories programmes appear naturally within this peg framework.  If we wish to discuss real Bayesian probabilities we could follow the linearly positive histories programme and take the real parts of our pegs and ensure a linearly positive condition \cite{Marlow06annals}.  Similarly, if we wish to discuss relative frequencies we could follow the consistent histories programme and take the real parts of our complex pegs and define a consistency condition stronger than linear positivity \cite{Dios04}.  Even so, we are still wary of invoking complex pegs because they are so alien to our usual notion of relative frequency.  Note, however, that we are not wholly uncomfortable as such complex pegs appear naturally in quantum theory---the generalised Berry phase is derived from such complex pegs \cite{AS01} and is an experimentally verifiable quantity (using ensembles of experiments).  As such, we can combine such phases and frequentist notions into one probabilistic entity using axioms that were outlined over 60 years ago by Cox \cite{Cox46}.

In the history of science we have been rather ambiguous about what the word `probability' means.  Some call relative frequencies `probabilities' even though they don't behave in the same manner as the term in common parlance.  Similarly we have called certain non-additive numbers in quantum theory `probabilities' even when they do not obey axioms of probability---nor axioms of relative frequency (which are necessarily additive).  We might call our complex pegs `probabilities' because at least they do obey rational probabilistic axioms, but perhaps we would make a similar category error or confuse the issue by doing so; hence, for the want of a better name, we have resorted to calling them `pegs'---at least it begins with a `p' so we don't need to change our notation.  So far we have two different kinds of pegs but there may be more.  We have objects that obey Cox's two axioms which are real; we might call these `round pegs'.  We also have these objects that obey analogues of Cox's two axioms, and our tentative third, which are complex.  Let us call these `square pegs'.  These names lead naturally to a playful, albeit unfortunately sardonic, metaphor for what we are trying to do.  A baby metaphor for science perhaps.  We are trying to find the right-shaped peg for the corresponding hole and we must reject the pegs that do not fit snugly.  When dealing with a histories algebra we argue that these complex pegs fit rather snugly.

\section{Entropy}

With a generalisation of probability to hand, we must also begin to discuss a generalisation of entropy---a complimentary concept that is often just as important as a good notion of probability.  Perhaps we don't have to search very far for such a generalisation.

First of all, how should a notion of entropy behave?  It should behave, in part, like a probability.  It should probably be a transitive or monotonic notion of preference in some sense \cite{Catich03}.  It should reflect the space of pegs in a natural way.  Hence the first na\"{\i}ve object to suggest is simply a generalisation of Shannon's entropy:

\begin{equation}
S[P(\unit_{\alpha} \vert I)] := -K_S \sum_{i=1}^n p(\alpha^i \vert I) \ln p(\alpha^i \vert I)
\label{entropytest}
\end{equation}

\noindent  where $P(\unit_{\alpha} \vert I) := \{p(\alpha_i \vert I) : i=1,2...N_{\alpha}\}$ and $K_S$ is a constant.  Does this object $S[\cdot]$ behave like an entropy should?  Does it, for example, obey the grouping property \cite{TS01}; a property that Shannon suggests is natural for any notion of entropy \cite{SWBOOK}.  Consider the complete---disjoint and exhaustive---set $\{\alpha^i\}$ split up into groups labeled by an integer $g$.  We could consider the peg-entropy (\ref{entropytest}) of the original set as split up into the peg-entropy of each of the groups and the peg-entropy as to which group $g=1,2,...,N_G$ one should use; this alternative way of looking at the peg-entropy (\ref{entropytest}) of the set should be equivalent to not splitting $\{\alpha^i\}$ into groups (this is the grouping property). How we split the entropy into groups should not make a difference.  We can split up the peg-entropy as follows:

\begin{equation}
S[P(\unit_{\alpha} \vert I)] := -K_S \sum_{g}^{N_G} \sum_{i \in g} p(\alpha^i \vert I) \ln p(\alpha^i \vert I).
\end{equation}

The complex peg we assign to a group $g$ is simply

\begin{equation}
p(g \vert I)= \sum_{i \in g} p(\alpha^i \vert I).
\end{equation}

\noindent So now we must ask ourselves whether

\begin{equation}
S[P(\unit_{\alpha} \vert I)] = S[P(\unit_G \vert I)] + \sum_g p(g \vert I) S_g
\end{equation}

\noindent where

\begin{equation}
S[P(\unit_G \vert I)] := -K_S \sum_{g}^{N_G} p(g \vert I) \ln p(g \vert I)
\end{equation}

\noindent and

\begin{equation}
S_g := -K_S\sum_{i \in g} p(\alpha^i \vert gI) \ln p(\alpha^i \vert gI).
\end{equation}

In order to work this out we need to work out what pegs we should assign to the histories $\{\alpha^i : i \in g \}$ upon the knowledge that the group $g$ is the correct group.  Thus we need a notion of conditioning.  We need to work out what conditional pegs we should use and whether it allows (\ref{entropytest}) to obey the grouping property.

Using Bayes' rule in our complex peg framework is quite interesting:

\begin{equation}
p(\alpha^i \vert g I) = \frac{p(g \vert \alpha^i I)p(\alpha^i)}{p(g \vert I)}.
\end{equation}

It is natural to assign $p(g \vert \alpha^i I) = 1$ if we know that $\alpha^i$ is in the group $g$ (we are normalising due to (\ref{COX2}) \emph{cf.} \cite{CoxBOOK}).  Hence we should assign

\begin{equation}
p(\alpha^i \vert gI) = \frac{p(\alpha^i \vert I)}{p(g \vert I)}
\end{equation}

\noindent to conditional grouping pegs.  Does this allow the grouping property to be satisfied?  Note that for $y,z \in \C$ we do not necessarily have that $\ln \frac{y}{z} = \ln y - \ln z$ because of different branches of the logarithm function; it only works if $-\pi  < (\arg(x) - \arg(y)) < \pi$.  Complex logarithms behave as follows:

\begin{equation}
\ln(r e^{i \theta}) = \ln r + i \theta
\end{equation}

\noindent  where we may choose the principle value of $\theta$.  Renaming the index $i$ with $j$ so as not to confuse it with imaginary components, it is therefore clear that:

\begin{equation}
\sum_j \ln \frac{p(\alpha^j \vert I)}{p(g \vert I)} = \sum_j (\ln \frac{\vert p(\alpha^j \vert I) \vert}{\vert p(g \vert I) \vert} + i \theta_j - i\theta_g)
\end{equation}

\noindent where $\theta_j = \arg[p(\alpha^j \vert I)]$ and $\theta_g = \arg[p(g \vert I)]$.

So, using the definition of the complex logarithm, we have that

\begin{equation}
-\sum_{j \in g} p(\alpha^j \vert I) \ln p(\alpha^j \vert I) = -p(g \vert I) \sum_{j \in g} \frac{p(\alpha^j \vert I)}{p(g \vert I)} \ln \frac{p(\alpha^j \vert I)}{p(g \vert I)} - p(g \vert I) \ln p(g \vert I).
\end{equation}

\noindent Thus we do have the grouping property for our test entropy functional (\ref{entropytest}) \emph{i.e.} we have that

\begin{equation}
S[P(\unit_{\alpha} \vert I)] = S[P(\unit_G \vert I)] + \sum_g p(g \vert I) S_g + 2 m \pi i
\label{grouping}
\end{equation}

\noindent where $m$ is an integer---(\ref{entropytest}) is satisfied as long as we identify the different branches of the logarithm.

What, other than the grouping property, should an entropy functional obey so as to be a useful definition of uncertainty or information?  According to Shannon \cite{SWBOOK}, $S[\cdot]$ should be continuous in the pegs.  When all the pegs are equal (and hence real $p_i=\frac{1}{n}$) then it should be a monotonic increasing function of $n$.  Hence $S[\cdot]$ should correspond to the standard Shannon entropy for the real subset of complex pegs.  Clearly (\ref{entropytest}) is very plausible as a generalisation of Shannon entropy that is apt for quantum histories.  But, like the Shannon entropy for real probabilities, is it possible to prove that we must use (\ref{entropytest}) because it is the only functional that fits the required desiderata up to some equivalence of functionals?  This we can't yet answer.

Are strong additivity and concavity also satisfied by this peg-entropy?  In order to find out we need to define a conditional peg-entropy; presumably this involves Bayes' rule which is satisfied by our complex pegs since the $\wedge$-operation is associative \cite{CoxBOOK}.  Lets define the conditional peg-entropy in an analogous way to how we define conditional Shannon entropies:

\begin{eqnarray}
S[P(\unit_{\alpha} \vert \unit_{\beta} I)] &:=& \sum_j p(\beta^j \vert I) S[P(\unit_{\alpha} \vert \beta^j I)]\\
&=& -K_S \sum_j p(\beta^j \vert I) \sum_i p(\alpha^i \vert \beta^j I) \ln p(\alpha^i \vert \beta^j I).
\end{eqnarray}

\noindent And thus we can check whether the following `strong additivity' condition is satisfied by $S[\cdot]$:

\begin{eqnarray}
S[P(\unit_{\alpha} \wedge \unit_{\beta} \vert I)] &=& S[P(\unit_{\alpha} \vert I)] + S[P(\unit_{\beta} \vert \unit_{\alpha} I)] \nonumber \\ &=& S[P(\unit_{\beta} \vert I)] + S[P(\unit_{\alpha} \vert \unit_{\beta} I)].
\label{strongadditivity2}
\end{eqnarray}

We can also check whether the following `concavity' condition is satisfied:

\begin{equation}
S[P(\unit_{\alpha} \vert I)] \geq S[P(\unit_{\alpha} \vert \unit_{\beta} I)].
\label{concavity2}
\end{equation}

\noindent Note that $P(\unit_{\alpha} \wedge \unit_{\beta} \vert I) := \{p(\alpha^i \wedge \beta^j \vert I): i=1,2,...n_{\alpha}  \mbox{ and } j=1,2,...n_{\beta}\}$.  So we can work out the LHS of (\ref{strongadditivity2}) to be

\begin{equation}
S[P(\unit_{\alpha} \wedge \unit_{\beta} \vert I)] =  - K_S \sum_{i}\sum_j p(\alpha^i \wedge \beta^j \vert I) \ln p(\alpha^i \wedge \beta^j \vert I).
\end{equation}

We can also work out the RHS (second decomposition) of (\ref{strongadditivity2}):

\begin{equation}
- K_S \sum_j p(\beta^j \vert I) \ln p(\beta^j \vert I) - K_S \sum_j p(\beta^j \vert I) \sum_i p(\alpha^i \vert \beta^j I) \ln p(\alpha^i \vert \beta^j I).
\end{equation}

Now, $p(\alpha^i \wedge \beta^j \vert I) = p(\alpha^i \vert \beta^j I)p(\beta^j \vert I)$ because Cox's axioms ensure that this is the case.    Note that $\sum_i p(\alpha^i \vert \beta^j I) = 1$ for each $j$ as long as $\alpha^i$ all commute with $\beta^j$.  Hence we can identify the LHS and RHS and thus (\ref{strongadditivity2}) is satisfied for sets of commuting histories.  Clearly it is natural that strong additivity applies for commuting histories because, in such cases, we can easily interpret the two sets of history propositions to be compatible.  If they do not commute then there is no \emph{a priori} reason we should demand strong additivity, just as there is no \emph{a priori} reason we should demand comparability (by the dubious axiom 0b) of probabilities in such cases.

In order to work out when (\ref{concavity2}) is also satisfied by our novel notion of entropy we would have to decide what partial order on the space $\C$ we ought to use.  Monoticity provides significant constraints upon what partial orders we can use and, as we argued above, it seems we should use the partial order illustrated in Fig.(\ref{partial1}) for pegs.  The partial order on the peg space will inform the partial order on the peg-entropy space (although perhaps scaled by the $K_S$ constant).  Concavity might then be satisfied, at least for a subset of histories.  Having shown that our tentative notion (\ref{entropytest}) obeys the grouping property (albeit identifying branches of the logarithm), it is a matter only of mathematical consequence whether our peg-entropy also obeys other convenient identities like strong additivity and concavity (these identities are not axioms \emph{per se}).  It is clear, then, that (\ref{entropytest}) is a plausible generalisation of entropy for quantum history theories but we have not yet proved whether all peg-entropies that obey the grouping property for complex pegs must be of this form.

\section{Conclusion}

In quantum theory we use Gleason's theorem to justify the probabilistic assignments we give to projection operators.  However, as soon as we begin to discuss more than one single projection operator---when we begin to discuss history propositions---we have to postulate a notion of state collapse in order to define probabilities.  However, such postulated probabilities are non-additive and many problems or issues arise because of this.  From a Bayesian perspective it is even dubious to call such things `probabilities' because they are non-additive and thus alien to our normal notion of probability \cite{Anast05}.  Problems with nonlocality also arise by discussing propositions that involve two or more times (in a given frame of reference).  Hence it is natural to tackle this problem head-on and define a propositional space that includes multi-time propositions.  Since we do not want to give any causal bias to the peg or probability theory that we use \cite{Leifer06} it seems prudent to put timelike and spacelike separated propositions on the same footing \cite{Hardy05}; hence we might na\"{\i}vely like to use tensor products to produce history propositions (this is the HPO algebra)\footnote{Of course, in full generality, one would prefer to use a fully relational propositional algebra.}.  Rather than postulate dubious notions of state collapse one can then \emph{derive} a monotonic peg for such history propositions, and one can do such a thing without getting into the problems of non-additivity and, tentatively, nonlocality.  There also exists a plausible generalisation of Shannon entropy for such pegs.  Of course such complex pegs are alien to our standard notion of probability.  However, our standard notion of probability is rather alien too; when you get down to it, what really does the term `probability' mean?  The interpretation of probabilities is clearly, historically speaking, a debatable issue and hence it is necessary to axiomatise and formalise a relational approach.  Such an approach will ensure that, even if we don't know with full clarity what such concepts mean, we will, in the least, not introduce functional relationships between pegs that we are not justified in introducing.

So we cannot yet give a clear answer to the question:  What are probabilities? One can, however, begin to answer another quite daunting question:  Why do we naturally find complex numbers in quantum theory?

\section*{Acknowledgements}

We would like to thank Lucien Hardy and Matthew Leifer for comments on a related talk and EPSRC for funding this work.

\end{document}